\documentclass[a4paper, conference, 10pt]{IEEEtran}
\usepackage{verbatim}
\usepackage{graphicx}
\usepackage{times}
\usepackage{epsfig}
\usepackage{epstopdf}
\usepackage{latexsym}
\usepackage{amsmath}
\usepackage{cite}
\usepackage{textcomp}
\usepackage{enumerate, flushend}
\usepackage{cancel}

\usepackage{authblk}\usepackage{algorithmic,algorithm}
\usepackage{subfig}
\begin{document}

\title{Performance Evaluation of Caching Policies in NDN - an ICN Architecture}
\author{Samar Shailendra, Senthilmurugan Sengottuvelan, Hemant Kumar Rath, Bighnaraj Panigrahi, Anantha Simha}
\affil{TCS Research \& Innovation, Bangalore, 560 100, India, \\ Email:\{s.samar, senthil.seng, hemant.rath, bighnaraj.panigrahi, anantha.simha\}@tcs.com}
\maketitle

\begin{abstract}
Information Centric Networking (ICN) advocates the philosophy of accessing the content independent of its location. Owing to this location independence in ICN, the routers en-route can be enabled to cache the content to serve the future requests for the same content locally. Several ICN architectures have been proposed in the literature along with various caching algorithms for caching and cache replacement at the routers en-route. The aim of this paper is to critically evaluate various caching policies using Named Data Networking (NDN), an ICN architecture proposed in literature. We have presented the performance comparison of different caching policies naming First In First Out (FIFO), Least Recently Used (LRU), and Universal Caching (UC) in two network models; Watts-Strogatz (WS) model (suitable for dense short link networks such as sensor networks) and Sprint topology (better suited for large Internet Service Provider (ISP) networks) using ndnSIM, an ns3 based discrete event simulator for NDN 
architecture. Our results indicate that UC outperforms other caching policies such as LRU and FIFO and makes UC a better alternative for both sensor networks and ISP networks.     
  
\end{abstract}


\section{Introduction}
Current Internet architecture was designed with a perspective of data exchange between two parties. Over the last decade, the nature of  content being accessed has changed dramatically. Today's user is more interested in accessing the content rather than the actual location of the same. Keeping this view in mind, Information Centric Networking (ICN) \cite{Xylomenos2013,Ahlgren2012} has been introduced as an alternate design for the Internet. The philosophy of ICN advocates accessing the content by its name and caching the same on the routers en-route between the source and the destination. The future requests for the content can be served from the nearby router with the same cached content. This requires a paradigm shift in the architecture of the Internet. Several architectures both overlay as well as clean state design have been proposed for the same \cite{Xylomenos2013}. 


In an ICN network, caching \cite{Tyson2013,Bari2012,Ahlgren2012} is one of the most crucial components for its performance. There are several caching policies being proposed in the literature with their relative merits and demerits. First In First Out (FIFO), Least Recently used (LRU), Least Frequently Used (LFU) are the basic caching policies being proposed for ICN caching. These caching policies are adopted from the Operating System's paging policies. There are other more sophisticated caching policies such as opportunistic caching where authors have suggested a cache aware routing algorithm  \cite{Sourlas2014}. Arianfar \textit{et al}. \cite{SomayaArianfar2010} have introduced a packet level caching where each packet is indexed with a preset probability and cached in the router. Caching has also been discussed in the context of Mobile Ad-hoc network (MANET) \cite{Kawadia2011,coacs} and web caching \cite{wang1999survey,Duarte2006,Che2001,Iyer2002}. Yin \textit{et al}. \cite{Yin2006} have introduced different variants of caching schemes based on path information and data information. In \cite{Panigrahi2015}, authors have proposed Universal Caching (UC) policy for ICN which accounts for various ICN networks related parameters such as distance from the source, reachability of the router, frequency of content access etc.

The desirable quality of any successful caching policy in ICN architecture is that it should be scalable and adaptable in variety of network topologies. In this paper, we have considered three caching policies (FIFO, LRU and UC) proposed in the literature for comparison. While there are several ICN architectures being proposed in the literature, in this work we have considered NDN architecture for our testing due to its popularity in the research community as well as an easy accessibility of ndnSIM, an ns3 based NDN simulator. ndnSIM currently supports FIFO, LRU, LFU etc. as the caching policies with FIFO as the default policy. We have further extended ndnSIM to support UC as the caching policy for ICN networks. 

In this work, we have implemented different network topologies namely Synthetic topology using Watts-Strogatz (WS) model \cite{watts1998collective} and Sprint topology \cite{spring2002measuring} in NDN architecture. These topologies are suitable to emulate the real world networks such as the dense network deployment and Internet Service Provider (ISP) network respectively. 

The rest of the paper is organized as follows. Section \ref{sec:uc} reproduces the details of the NDN architecture, Caching policies and brief note on UC implementation in ndnSIM. In Section \ref{sec:model}, we outline the details of the system model, such as network topology, and request distribution etc.. Section \ref{sec:perf} gives the performance comparison of different caching policies followed by the conclusion in the subsequent section.

\section{NDN architecture and Caching Policies}
\label{sec:uc}
In this section we provide an overview of the NDN architecture. We have used ndnSIM, an ns3 based simulator, to simulate the NDN architecture. ndnSIM is distributed with native support of FIFO and LRU caching policies. We have enhanced it to support UC caching policy. A note on NDN architecture and implementation of UC algorithm in ndnSIM has also been provided in the subsequent subsections.

\subsection{NDN Architecture}
Named Data Networking (NDN) is a receiver-driven network where user has to request data by sending \textit{Interest} packet with the name of the desired content. The \textit{Interest} packets are routed to the data source based on the name in the packet. The \textit{Data} packet whose name matches the name in the \textit{Interest} packet is served to the user through the reverse path. In NDN, routers contain three tables namely Pending Interest Table (PIT), Content Store (CS), and Forwarding Interest Base (FIB), which help in forwarding, caching and serving of contents etc. 

\begin{figure}[ht!]
\centering
\includegraphics[scale=0.6]{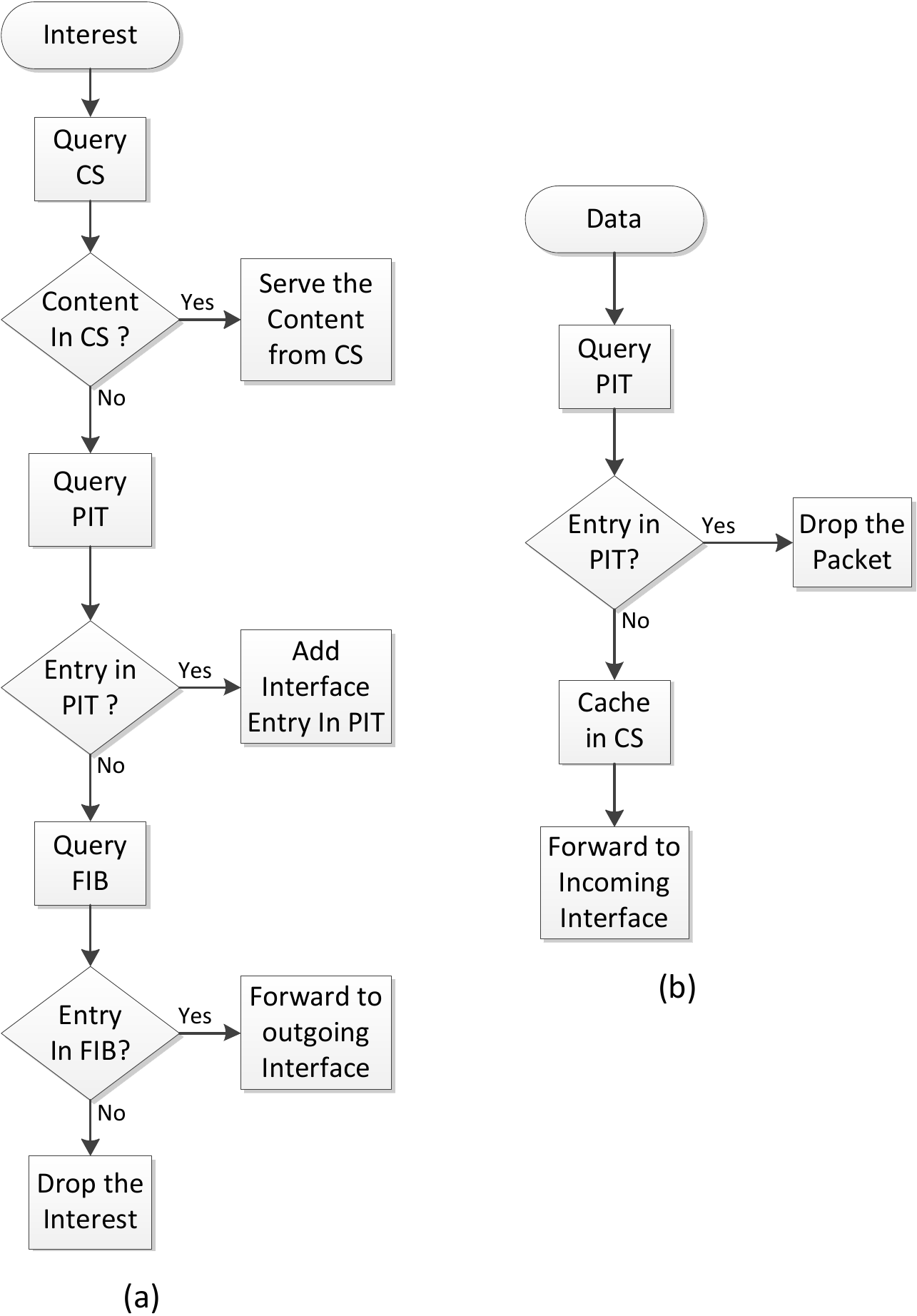}
\caption{(a) Interest flow and (b) Data flow in NDN Architecture}
\label{fig:NDN}
\end{figure}

When an \textit{Interest} packet arrives at a router, the name of the desired content is queried in router's CS. If the content is found in CS, the \textit{Interest} is served by sending the \textit{Data} packet through the incoming interface. If the content is not found, then the router checks its PIT table for an entry of the same content. If a duplicate entry is found, the ingress interface is added to the existing entry and the current \textit{Interest} packet is dropped. Otherwise, a new entry is created in PIT and the \textit{Interest} packet is forwarded to the next outgoing interface based on FIB table. The \textit{Data} packets are served through reverse path by looking at entries in PIT and forwarding to all interfaces in the matching PIT entry. These \textit{Interest} and \textit{Data} flows are summarized in Fig. \ref{fig:NDN} 

The cache size of CS is only a small fraction of the total number of contents available in the network. Moreover, the popularity of all contents are not same and also the popularity varies with time. Hence, the caching algorithm plays a vital role in improving the performance metrics such as cache hit of a network.

\subsection{Caching Policies}
In NDN, the content is being replaced from its Content Store (CS) based upon the underlying caching policy. FIFO is the default caching policy in NDN. In FIFO the content that has arrived first in the CS, the oldest content in the cache, is replaced by the newly arriving content. LRU is another caching policy natively supported by ndnSIM. LRU replaces the data based upon the time of usage of the content i.e. the one which is not being used for the longest time is replaced. In the literature, UC \cite{Panigrahi2015} is being proposed as an alternate policy designed for information centric networks. UC assigns a Content Metric (CM) to any incoming content and replaces the content based upon its CM. We have implemented UC in ndnSIM to compare it with other policies. Some of the implementation notes are highlighted below.

\subsection{Implementation of Universal Caching in ndnSIM}
\begin{figure}[ht!]
\centering
\includegraphics[scale=0.5]{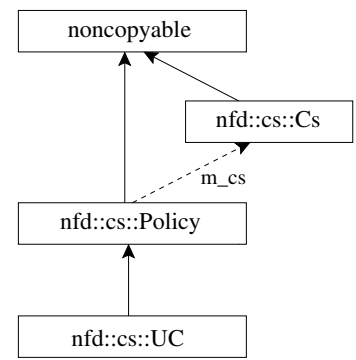}
\caption{Collaboration diagram for UC in ndnSIM}
\label{fig:ndn_uc}
\end{figure}

In ndnSIM, caching modules are implemented in NDN Forwarding Daemon (NFD) module. We have used NFD based CS module to implement Universal Caching algorithm. We have inherited new class named `UC' from already existing abstract class \emph{nfd::cs::Policy}. The newly inherited class also collaborates with Content Store class \emph{nfd::cs::CS} of NFD daemon. The inheritance and collaboration diagram for \emph{nfd::cs::UC} is shown in Fig. \ref{fig:ndn_uc}. ndnSIM uses Boost Multi-index Containers library which provides multi level indexes to create indexing of the contents for different caching policies. 

\section{Performance Evaluation Model}
\label{sec:model}

In this section we present the details about the network topologies and the content request pattern used for our simulations. It is very important to test the performance of any algorithm in large real world network topologies. We have used ndnSIM to simulate the large scale ICN networks. 

\subsection{Network Topologies} 
For our evaluation, we have considered two different models for network topologies to compare the performance of different caching policies in NDN. 

\subsubsection{\textbf{Synthetic Topology}}
We have generated synthetic topologies using Watts-Strogatz (WS) model. WS model \cite{watts1998collective} is a random graph for high density network with smaller path lengths which makes it suitable for dense networks such as sensor network where a lot of sensors are deployed within small area. Steps to generate Synthetic topology are 

\begin{itemize}
 \item Let $C(n,k)$ be the graph with nodes \{$a_i$, 0 $\le$ $i$ $\le$ ($n$-1)\} and edges \{($i$,$j$), 0 $\le$ $i$,$j$ $\le$ ($n$-1) and $\vert$ $i$ - $j$ $\vert$ $\le$ $k$/2\},
 \item Graph $WS(n,k,p)$ is obtained by replacing each edge with probability $p$ in $C(n,k)$.
\end{itemize}

In synthetic topology, $n$ is the number of routers in the network. Each router may have a number of users attached with it and requesting for content. Each router in the network is connected to at least $k$ other routers where the network graph of these routers is generated using the steps described above.    

The Stanford Network Analysis Platform (SNAP) \cite{snap2013} tool is used to generate WS based random topologies with fixed number of nodes. A python script is developed to convert the output of SNAP to a format that can be read by ndnSIM's annotated topology reader. 

\subsubsection{\textbf{Sprint Topology}}
We have also evaluated the performance of caching algorithms on realistic ISP topologies of existing Internet. The ISP network has multiple Points of Presence (POPs) where the ISP houses the collection of routers. All these POPs are linked together by a \textit{backbone} or \textit{core} router. Each POPs have \textit{access} router that connects with backbone routers and neighbouring POPs. The evaluation of caching algorithms on router-level topologies, which has 400 routers on an average in ndnSIM, is quite time consuming. Thus, we have used POP-level topology to mimic the large ISP networks. In particular, the Sprint topology \cite{spring2002measuring} with 52 POP level routers is used in our simulation. In this setup, all the routers are assumed to receive the data requests at random time instances during the course of simulation.  

%

\subsection{Request Arrival distribution}
In traditional networks, the content arrival is considered to be following Poisson Distribution. However, in modern day traffic, it has been observed that the contents arriving at any node have higher correlation. The distribution of such content arrival in the Internet is modeled using Heavy Tail Distribution. It has also been observed in the literature that the request arrivals at any router can be better approximated using Mandelbrot-Zipf (MZipf) distribution \cite{pentikousis2014, Hefeeda2008}. According to MZipf distribution, the probability of accessing an object at rank $r$ is: 

\begin{equation}
p(r)=\frac{H}{(r+q)^\alpha},
\label{eq:zipf}
\end{equation}
where, 

\begin{equation}
H=\sum_{c=1}^{N} \frac{1}{(c+q)^\alpha},
\end{equation}

\noindent and $\alpha$ is the \textit{skewness factor} which controls the slope of the curve, while $q \ge 0$ is known as the \textit{plateau factor} and decides the flatness of the curve \cite{Saleh2006}. In our system model, we have considered that the requests arrivals follow MZipf distribution.

 \begin{figure*}
 \centering
\subfloat[With Varying Cache Sizes]
{\includegraphics[width=0.4\linewidth]{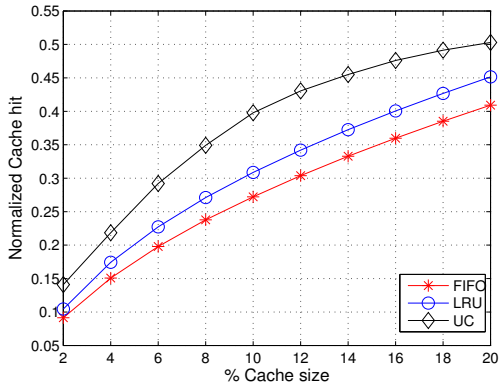}}
\subfloat[Different realization of Topologies]
{\includegraphics[width=0.4\linewidth]{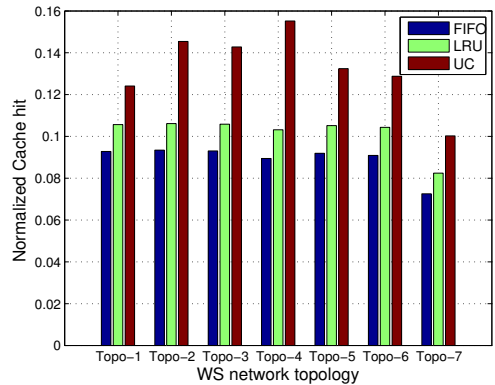}}
\caption{Normalized Cache hits for Synthetic Topology (a) with varying cache size (b) with different realizations }
\label{fig:hit_synth}
\end{figure*}

 \begin{figure*}
 \centering
\subfloat[With Varying Cache Sizes]
{\includegraphics[width=0.4\linewidth]{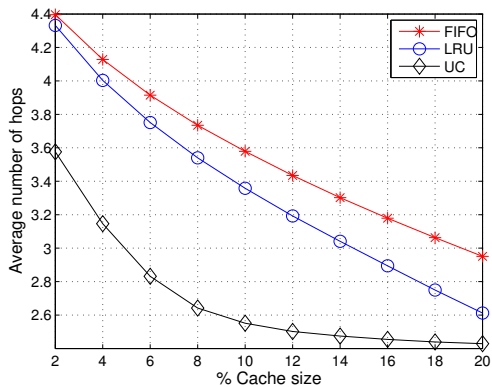}}
\subfloat[Different realization of Topologies]
{\includegraphics[width=0.4\linewidth]{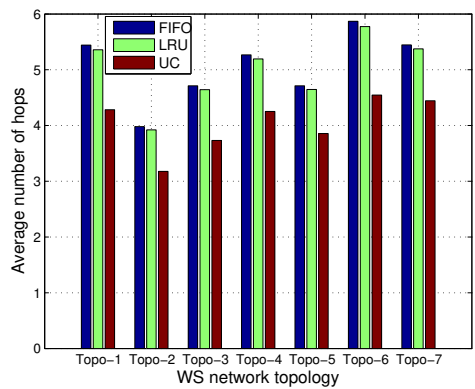}}
\caption{Average Hop count for Synthetic Topology (a) with varying cache size (b) with different realizations}
\label{fig:hop_synth}
\end{figure*}

\section{Performance Results}
\label{sec:perf}
In this section, we present the performance comparison of different caching policies for synthetic topology and Sprint topology. All the users are considered to generate the requests using MZipf distribution.

For our comparison, we consider the normalized cache hits (cache hits with respect to the total number of unique content requests at the router) in the network and average number of hops required to get the requested content as the performance parameters. We have obtained the variation of these parameters with respect to the percentage cache size. The percentage cache size is the cache size at any router in comparison to the unique content being requested at the router. Note that in these simulations hop count is also a direct measure of the latency because all paths are of identical characteristics. The requests are generated at a rate of 1000 requests per second in these simulations. Each ICN chunk/packet is considered to be of uniform size of 1KB and all result values obtained are withing the 95\% of the confidence interval.

\subsubsection{Synthetic Topology}
For the performance comparison of the caching policies, the requests are generated with MZipf distribution with parameter $q$=0.7 and $\alpha$=0.7 \cite{pentikousis2014}. Different realizations of topology using WS model are being generated with parameters $n$=100, $k$=2 and $p$=0.1. 

The results obtained in Fig. \ref{fig:hit_synth}(a) clearly show that normalized cache hits for UC are always more than the FIFO and LRU caching policies. It is interesting to note that initially UC provides better incentive with respect to the increase in the cache size. However, beyond a point there is no proportional incentive in increasing the cache size because most of the popular content has already been stored in the caches. Moreover, on increasing the cache size further, the performance of all the three caching policies are similar because most of the requested content is stored in the cache. There is also a significant reduction in the average number of hops required to get the contents from the router (Fig. \ref{fig:hop_synth}(a)).

We have also obtained the performance comparison of the three caching policies for different random realizations of the Synthetic topology (Topo-1 to Topo-7) in Fig. \ref{fig:hit_synth}(b) and Fig. \ref{fig:hop_synth}(b). For each realization of the synthetic topology, UC performs consistently better than other policies in every situation. These results indicate that UC is a better choice for dense network deployments for an ICN network. 

\subsubsection{Sprint Topology}
\begin{figure*}
\subfloat[Zipf parameters $q$=0.7, $\alpha$=0.7]
{\includegraphics[width=0.32\linewidth]{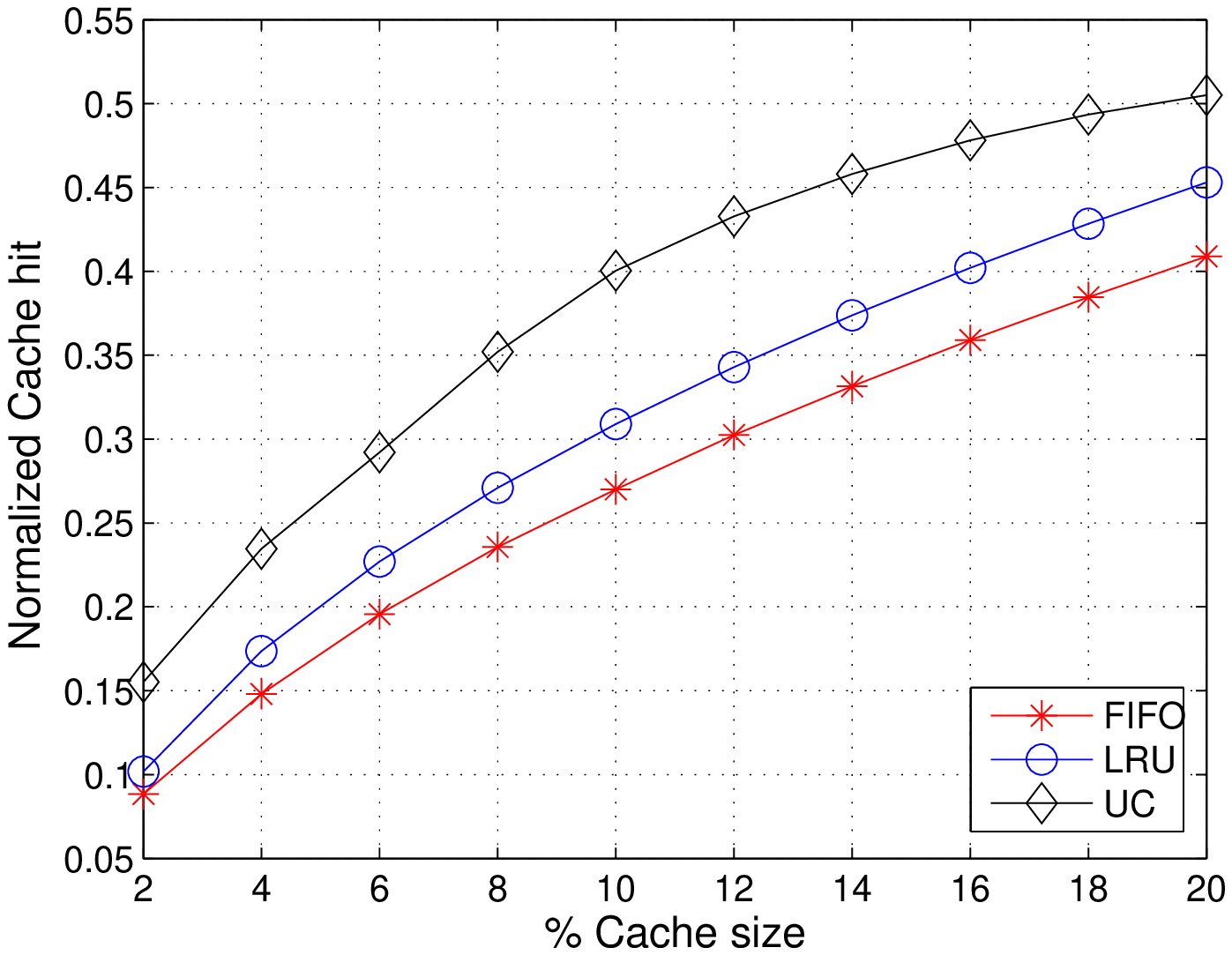}}
\subfloat[Zipf parameters $q$=5, $\alpha$=0.65]
{\includegraphics[width=0.32\linewidth]{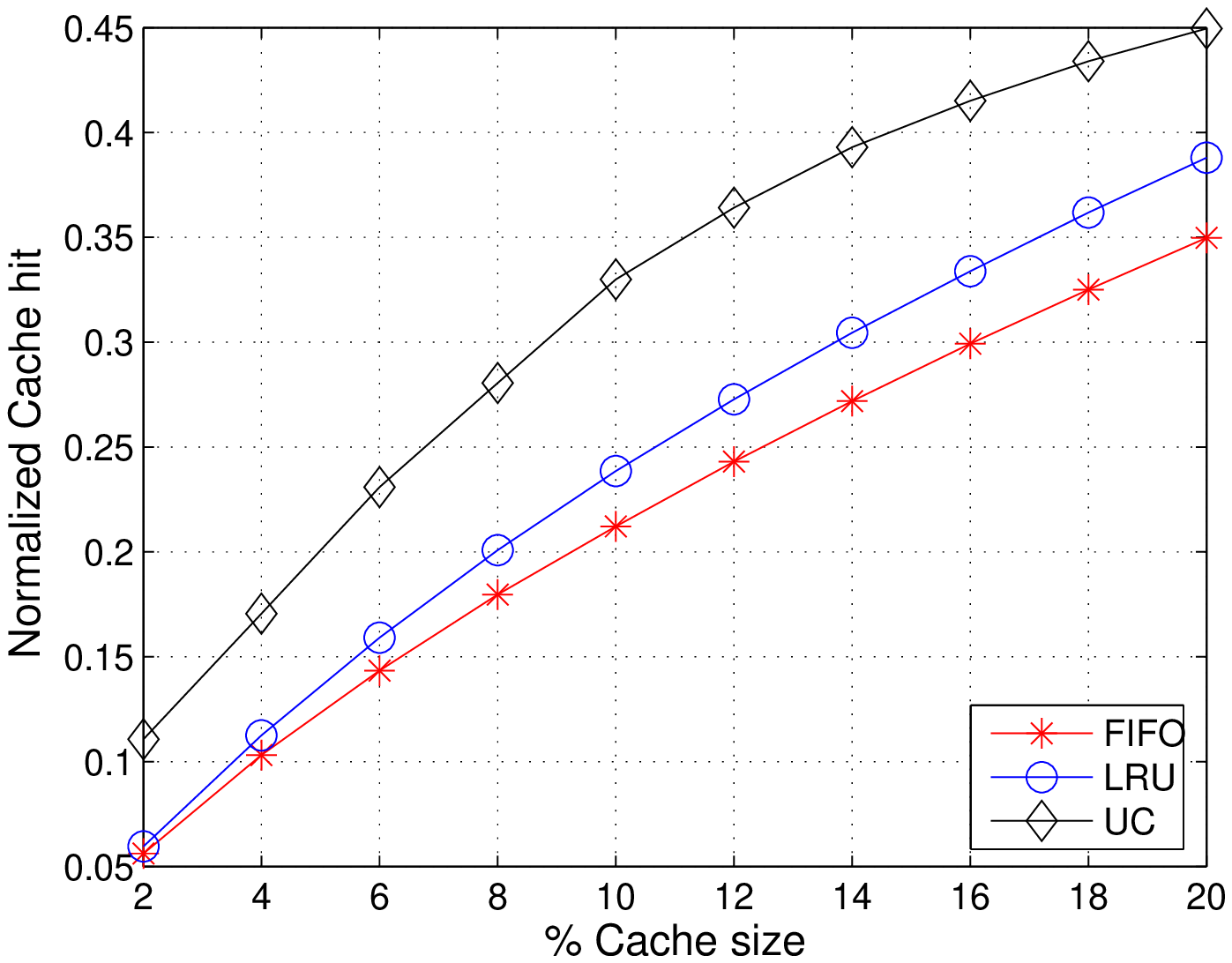}}
\subfloat[Zipf parameters $q$=55, $\alpha$=0.6]
{\includegraphics[width=0.32\linewidth]{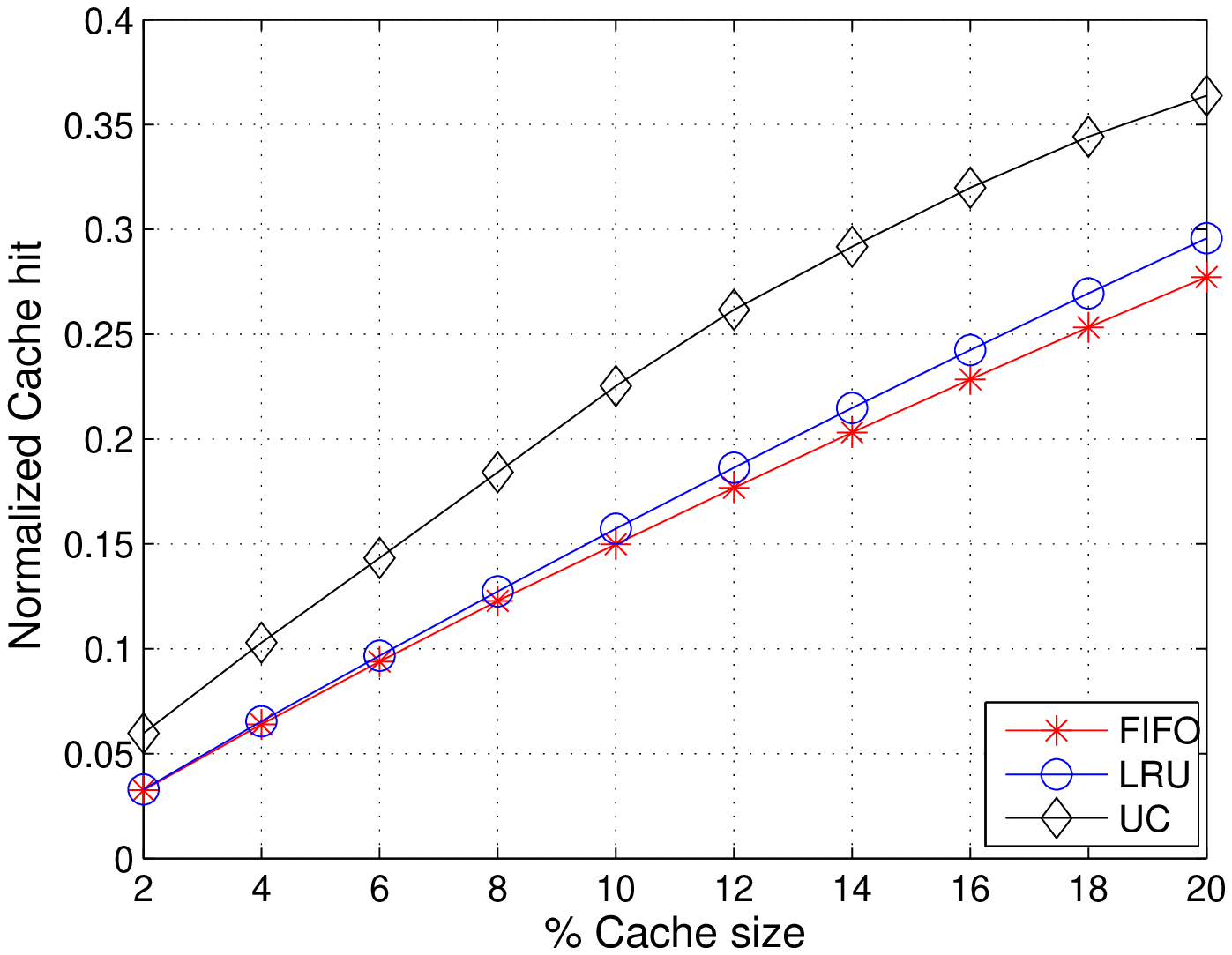}}
\caption{Normalized Cache hits with varying cache size for Sprint Topology }
\label{fig:hit_spr}
\end{figure*}

\begin{figure*}
\subfloat[Zipf parameters $q$=0.7, $\alpha$=0.7]
{\includegraphics[width=0.32\linewidth]{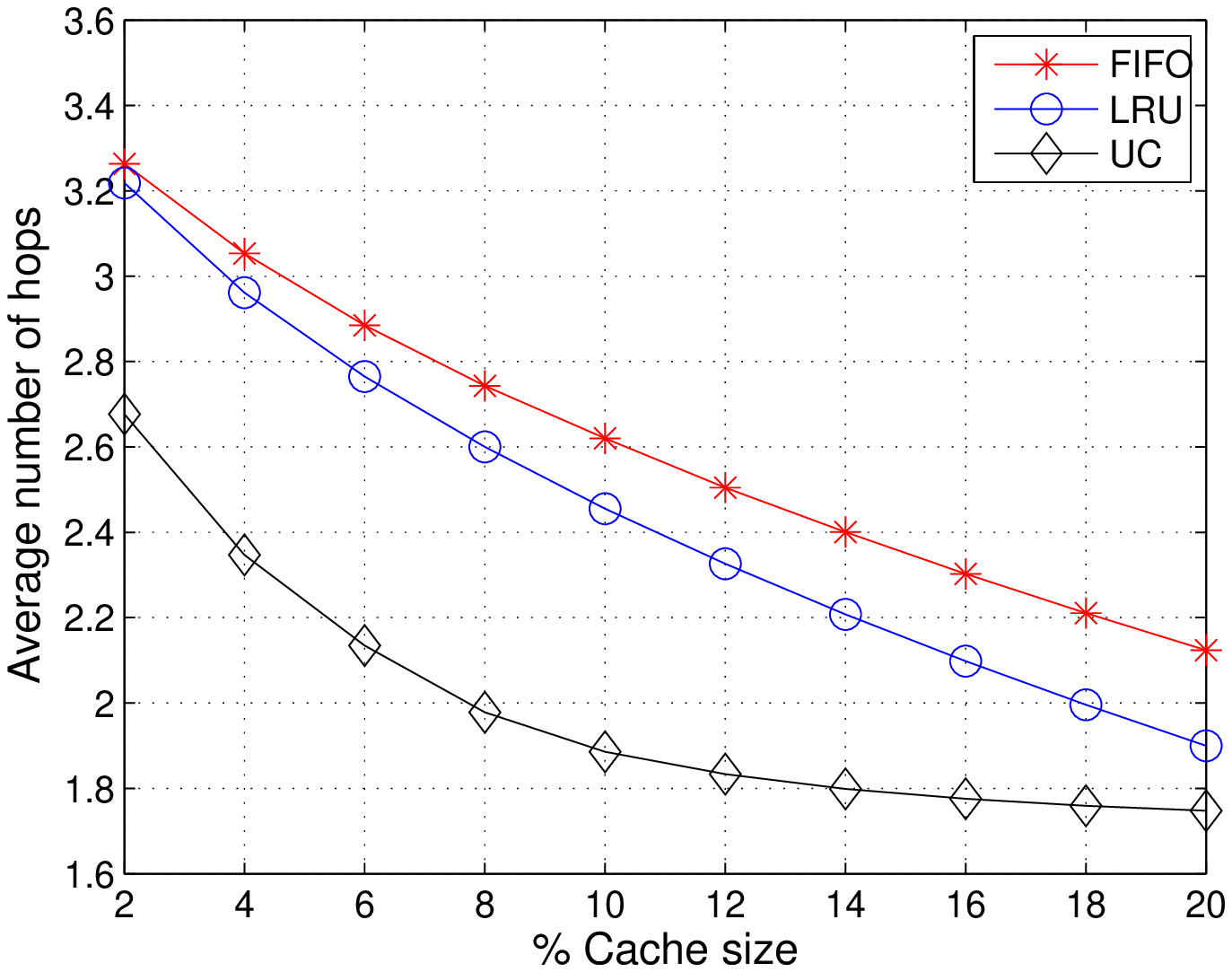}}
\subfloat[Zipf parameters $q$=5, $\alpha$=0.65]
{\includegraphics[width=0.32\linewidth]{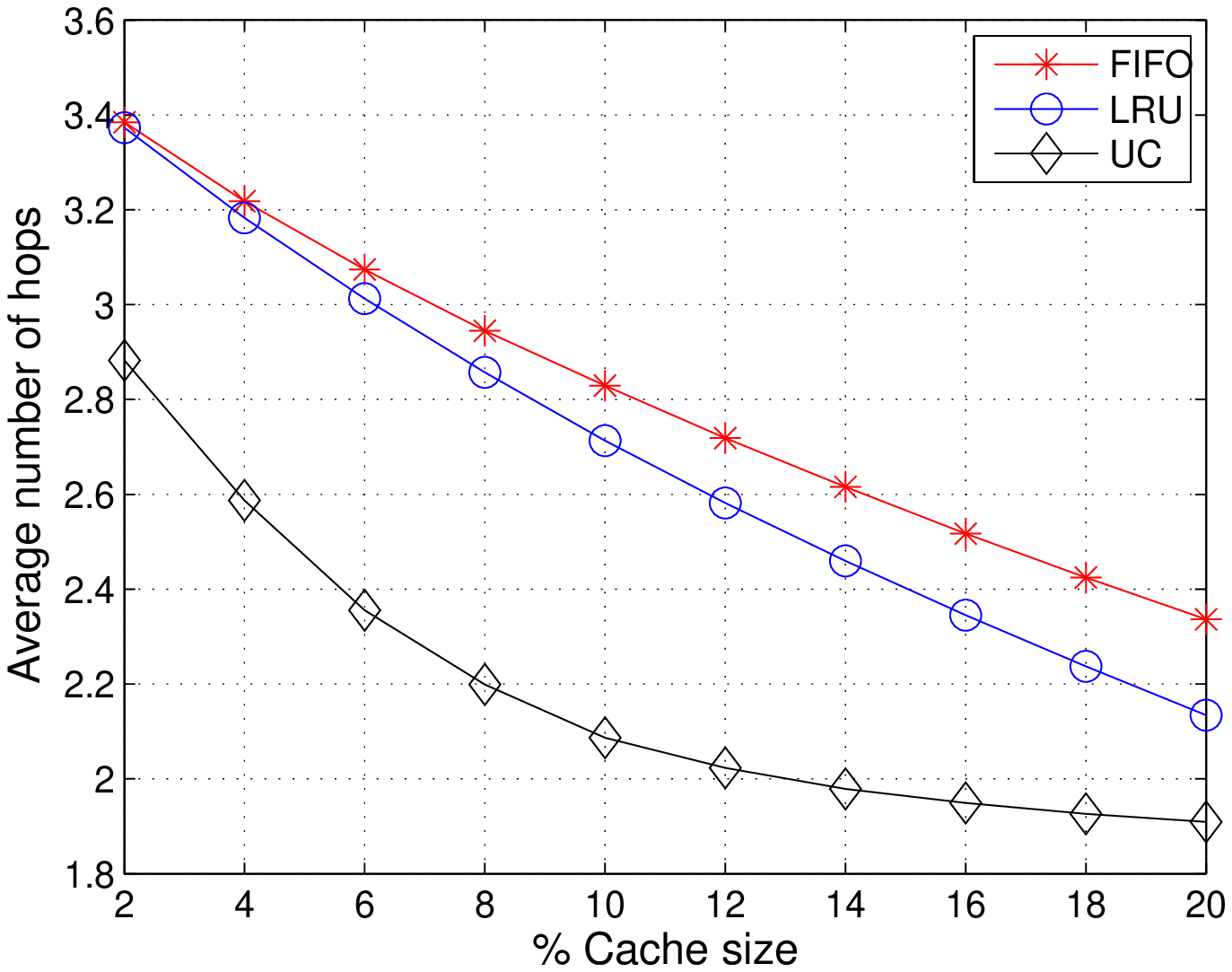}}
\subfloat[Zipf parameters $q$=55, $\alpha$=0.6]
{\includegraphics[width=0.32\linewidth]{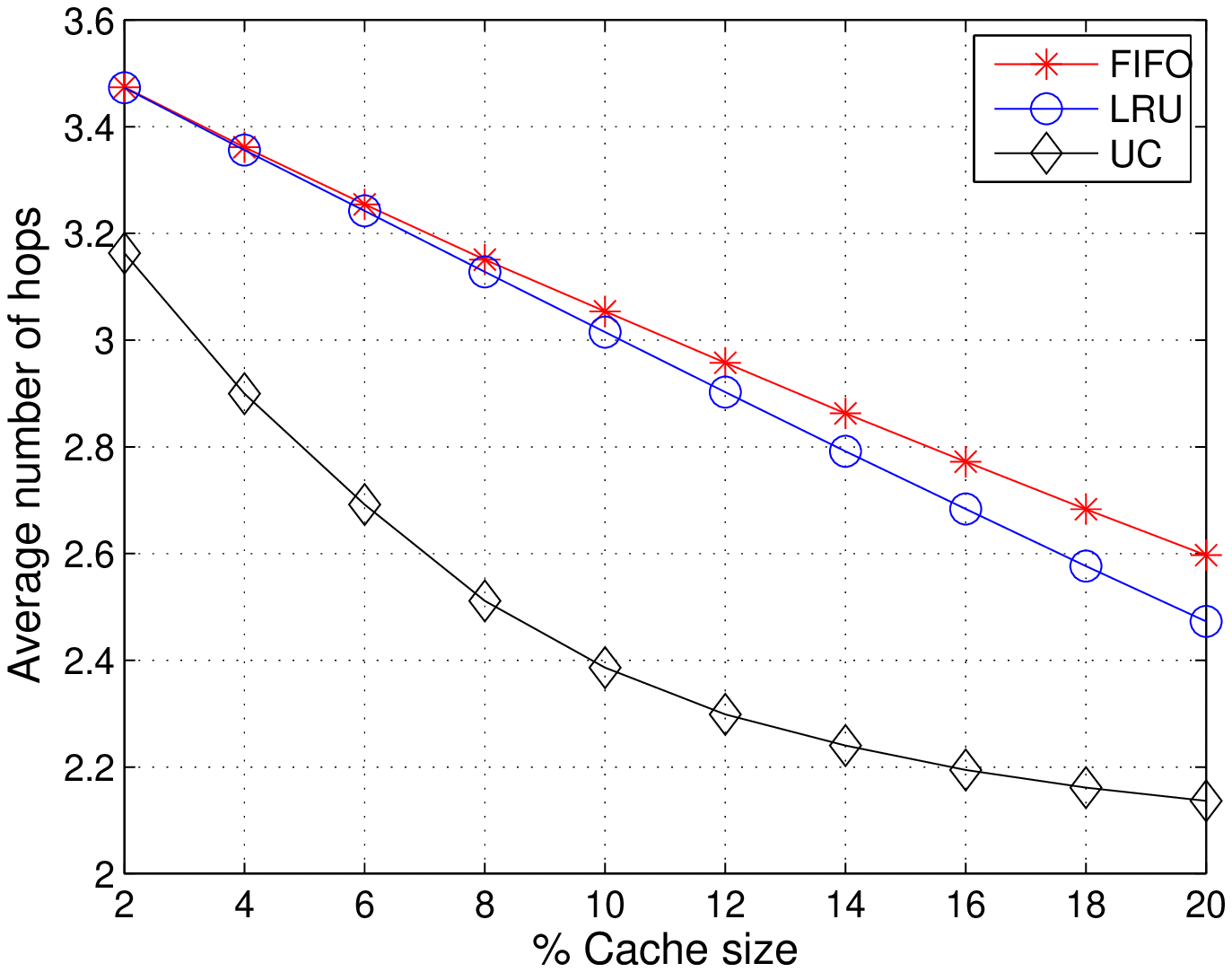}}
\caption{Average number of hops with varying cache size for Sprint Topology }
\label{fig:hop_spr}
\end{figure*}

Sprint topology mimics the large ISP networks more accurately. Content distribution in the Internet follows the heavy tailed distribution. Here, we have compared the caching policies for different values of MZipf parameters. 
Fig. \ref{fig:hit_spr} shows the normalized cache hits for different caching policies for different MZipf parameters. We have also plotted the average number of hops required to download the content for different caching policies for different MZipf parameters (Fig. \ref{fig:hop_spr}).

For the MZipf distribution as the content popularity increases, the normalized cache hits as well the average hop count improves. This is expected because lesser number of unique contents are accessed more frequently with increase in content popularity. It is also interesting to note that UC not only provides better performance in comparison to other caching policies but also provides better utilization of the caching resources at the routers in the network.

\section{Conclusion}
\label{sec:concl}
In this paper, we have implemented Universal Caching in NDN architecture using ndnSIM and presented the performance comparison with LRU and FIFO caching policies. We have also simulated synthetic and Sprint topology models in ndnSIM which are able to mimic dense network deployment and large ISP networks, respectively. The performance results of the caching policies in these network topologies show that Universal Caching performs better than other caching policies for ICN networks. The results also indicate that UC provides better incentive in terms of performance improvement with an increase in caching resources. More experiments can be conducted in future to test the performance of UC algorithm for other network architectures such as Fog computing and also in comparison with other caching policies proposed in literature.

 \bibliographystyle{ieeetr}
 \bibliography{ICN_noLink.bib}

\end{document}